\documentclass[12pt,british]{article}
\usepackage[T1]{fontenc}
\usepackage[latin9]{inputenc}
\usepackage{geometry}
\usepackage{amsmath}
\usepackage{amsthm}
\usepackage{setspace}
\usepackage[authoryear]{natbib}
\makeatletter
\theoremstyle{plain}
\newtheorem{lem}{\protect\lemmaname}
\theoremstyle{plain}
\newtheorem{prop}{\protect\propositionname}
\theoremstyle{plain}
\newtheorem{cor}{\protect\corollaryname}

\usepackage{hyperref}
\usepackage{tikz}
\usepackage{pgfplots}

\makeatother

\usepackage{babel}
\providecommand{\corollaryname}{Corollary}
\providecommand{\lemmaname}{Lemma}
\providecommand{\propositionname}{Proposition}

\begin{document}
\title{Voting behind the Veil of Ignorance\thanks{I thank V Bhaskar for advice, as well as Steven Callander, Antoine Loeper, and Francesco Squintani, for helpful comments. Financial support from MICIN/AEI/10.13039/501100011033 grants CEX2021-001181-M, PID2020-118022GB-I00, and RYC2021-032163-I; and Comunidad de Madrid grants EPUC3M11 (V PRICIT) and H2019/HUM-5891 is gratefully acknowledged.}}
\author{Boris Ginzburg\thanks{Department of Economics, Universidad Carlos III de Madrid, Calle Madrid 126, 28903 Getafe (Madrid), Spain. Email: bginzbur@eco.uc3m.es.}}
\maketitle
\begin{abstract}
A committee consisting of two factions is considering a project whose distributive consequences are unknown. This uncertainty can be resolved at some unknown future time. By delaying approval, the committee can gradually learn which faction benefits from the project. Because support of both factions is required for approval, it can only happen when there is sufficient amount of uncertainty about the identities of winners and losers. I show that in many situations, a project is more likely to be approved if it gives a lower payoff to everyone. The probability of approval and expected payoffs of both factions are higher if the project is ex ante less likely to benefit the faction that tends to receive good news faster. Equilibrium amount of learning is excessive, and a deadline on adopting the project is often optimal.


\end{abstract}
\newpage

\section{Introduction}

Distributive consequences of collective decisions are often unknown, and are revealed at uncertain time. Consider, for example, a national legislature that is deciding whether to fund building a railway line that would run through two regions. If the railway is built, the improvement in transport connections will encourage a major firm to open a factory in one of these regions. Members of the legislature from both regions need to agree for the railway to be built,  both regions would have to share the cost of building it, but only the region in which the factory will be opened will reap the benefits. The firm has not yet decided in which region it will build the factory, but it will announce the prospective location at an unknown future date. At any moment, the legislature can vote in favour of building the railway. Or it can vote to delay the decision, and as it waits, it updates its belief about the prospective location.

This situation has several distinguishing features: (i) broad support is required for the proposal to be adopted; (ii) the identities of winners and losers are initially unknown; (iii) this information will be revealed publicly at an unknown time; (iv) members of the decision-making body can collectively decide to delay adoption of the proposal. Such situations are common in political and economic settings, as discussed below. This paper explores their implications for voting outcomes, payoffs, and optimal decision rules.

In more detail, the paper models a committee consisting of two factions, called $A$ and $B$. The committee needs to decide whether to implement a project at a cost to both factions. The project can be of two types. One type of the project brings a positive payoff to faction $A$, while the other -- to faction $B$. The type is initially unknown, and there is a common belief about it. Time is continuous. At any point in time, the committee can either approve the project, which ends the game, or continue waiting. As long as the committee is waiting, at any point a public signal can arrive and reveal the project's type. The type that favours faction $A$ is more likely to be revealed sooner -- hence, as long as no signal has arrived, the probability that the project benefits faction $A$ is decreasing. 

Approval of the project requires support of both factions.\footnote{If one faction has sufficient weight to be able to force approval on its own, the problem becomes equivalent to a standard model of individual experimentation as in \citet{keller2005strategic}.} Because of this, once the type of the project is revealed, the faction that does not gain from the project will vote against it forever, blocking its approval. However, the project can be approved when the common belief about the type is such that each faction is sufficiently likely to benefit from the project. Thus, as in the famous thought experiment by \citet{rawls1971}, agreement can be found when the identities of the winner and the loser are hidden behind the veil of ignorance.

The point at which approval happens is determined by the asymmetry in the arrival of information. Because the belief is only becoming worse for faction $A$, it does not gain from learning. Instead it simply votes for the project until its expected payoff from approving it becomes negative. Faction $B$, on the other hand, gains from any delay, because this reduces the chance that a project from which it does not gain is approved. However, it knows that after a certain point, faction $A$ will cease to support the project. To avoid this, faction $B$ may switch its decision at that point, and vote for the project.

This logic leads to three main results. First, for a significant range of parameter values, the project is more likely to be approved whenever its cost is higher. Hence, a project that is ex ante and ex post\emph{ worse for everyone} has a higher probability of being implemented than a more efficient project. In terms of the above example, a railway is more likely to be built if it costs more. The reason is that, when the cost of the project is higher, faction $A$ stops supporting it earlier. This forces faction $B$ to also switch its decision earlier, or else the project would never be implemented. Hence, the two factions agree to approve the project at an earlier time. This leaves less time for the type to be revealed (which would make approval impossible). 

Second, even though the decision rule is symmetric -- both factions need to support the project in order for it to be implemented -- the initial belief does not affect the chance of approval symmetrically. Instead, a project that is ex ante less likely to benefit faction $A$ has a higher probability of being approved. Moreover, such a project is better for \emph{both} factions. The reason is that with such a project, there is less time for the belief to reach a point when faction $A$ switches its decision. This leaves less time for a signal to arrive, increasing the chance of approval, which makes both factions better off. Returning to the initial example, suppose that there are several potential routes for the railway. A route that gives region $A$ a more convenient connection makes it more likely that the firm will choose region $A$ for its factory once the railway is built. Then if representatives of region $A$ are the ones proposing the route, they will put forward one that is less convenient for their own region.

Third, I show that at the equilibrium, the committee always acquires weakly more, and sometimes strictly more, information than the optimal amount -- that is, delays the decision for too long. Because of this, it is often socially optimal to impose a deadline, which would force the committee to approve the project earlier, limiting information acquisition.


The setting described by the model relates to several types of collective decisions. The next few paragraphs discuss some of the examples.

\paragraph{Infrastructural investment.}

With depressing frequency, transport infrastructure projects in many countries suffer from cost overruns and fail to achieve their objectives.\footnote{See \citet{flyvbjerg2003common,cavalieri2019tales,denicol2020causes}. For instance, \citet{brooks2023infrastructure} show that in the United States, the costs of highway construction haves increased more than threefold between 1960s and 1980s, even though increases in material and labour costs cannot explain these changes.} Why does this happen? One feature of such projects is that they bring uneven benefits across regions, as improvements to transport can lead to reallocation of workers and firms between them \citep{qin2017no,heuermann2019effect,fretz2022highways}. At the same time, the distribution of benefits is often difficult to predict in advance in each specific case.\footnote{See \citet{cheng2022socioeconomic} for an overview of studies aiming to assess distributional impact of railway construction.} One way to learn them is to delay the decision until a comprehensive study or a pilot stage is completed, which reveals the distributive consequences of the project. The government that funds the project can either do this, or commit to the project from the start, without waiting to gather this information. The model suggests that in such situations, the project is more likely to receive broad support if it is less efficient -- that is, if it carries a higher cost relative to the sum of benefits.

\paragraph{Environmental regulation.}

Research suggests that instruments such as Pigouvian taxes on emissions may, at least in many situations, be more efficient in mitigating climate change than emission quotas or other command-and-control instruments \citep{hepburn2006regulation,stiglitz2019addressing}. Yet policy-makers choose command-and-control interventions significantly more often \citep{guo2021can}. One reason may be that the distributional effects of environmental regulation are uncertain. Firms that focus on green technologies stand to gain from such policies, but the identities of specific winners depend on which of several emerging technologies turns out to be more promising. This is initially unknown, and is only revealed once a technological breakthrough arrives. The time at which it happens is likewise unknown. The model suggests that in such situations, it may be easier to find consensus for introducing a less efficient policy instrument.

\paragraph{Pharmaceutical industry and intellectual property rights.}

In the United States, pharmaceutical firms expend considerable resources on lobbying in order to convince legislators to extend patent terms.\footnote{\citet{scherer1993pricing} and \citet{chu2008special} describe lobbying effort by the US pharmaceutical firms to extend intellectual property rights for new drugs.} If successful, such lobbying brings significant benefits to innovating firms. However, lobbying requires joint effort on the part of pharmaceutical companies. When is such coordination likely to succeed? Consider two firms engaged in a race to develop a new drug. If they jointly decide to lobby for extending intellectual property rights, the firm that eventually wins the race will reap the benefits of extended patent protection. The identity of the winner is revealed when one of the firms achieves a breakthrough, but the time at which it happens is uncertain. The above results suggest that such joint lobbying may be more likely to succeed when the cost of lobbying efforts, relative to the value of the market for the new drug, is high.

\subsection{Related literature.}

The paper contributes to the literature on collective deliberation, in which committees vote to acquire information about a payoff-relevant state prior to making a decision. In \citet{chan2018deliberating}, a committee is deciding on whether to approve one of two alternatives or continue gathering information about a state of the world. All members want to select the alternative that matches the state, although preference intensities differ. The paper analyses the impact of voting rules, preference intensities, and impatience on the ability of the committee to make the correct decision. \citet{anesi2018policy} study policy experimentation by a committee which, at each stage, votes on a tax rate, on a redistribution scheme, as well as choosing between a risky reform and a safe alternative; selecting the reform enables learning whether it is good or bad for all members.\footnote{A number of papers also study policy experimentation by multiple alternating principals, but without voting. See, for example, \citet{callander2014preemptive}.} \citet{anesi2022deciding} examine the impact of rules that allow costly deliberation to stop. In their model, at each stage, the committee can vote to stop information acquisition; doing so enables it to make a vote on approving or rejecting the reform. The paper shows how such deliberation rules can bring Pareto-inefficient outcomes.


Crucially, in this literature all voters prefer the same alternative in a given state -- a classic example is a jury whose members agree that the defendant should be convicted if he is guilty, and acquitted if he is innocent. Because of this, under full information the committee would always make a socially optimal decision, and members would always prefer acquiring more information if doing so was costless. A key difference of my model is that factions have opposing preferences.\footnote{For a model of voting under adversarial preferences in a setting without collective learning, see \citet{kim2007swing}.} Consequently, when information is fully revealed (and, more generally, when the common belief is close to zero or one) the project is always rejected even if it is socially optimal to adopt it. Instead, the project can only be adopted under incomplete information. Hence, the committee chooses to stop acquiring information even when members are infinitely patient. The fact that information prevents approval implies the key results, such as the optimality of the deadline (which limits information acquisition), as well as the fact that more costly projects are more likely to be approved (as they induce less learning).

The closest paper that similarly studies information acquisition by a committee with potentially conflicting preferences is \citet{ginzburg2019collective}. In it, a committee decides whether to learn a state of the world before choosing to accept or reject a proposal. However, in \citeauthor{ginzburg2019collective}, learning the state is a one-shot decision. This corresponds to a setting in which the state is revealed at a predetermined time, and the committee can choose whether to delay the decision until then. Here, on the other hand, the time at which a signal arrives is uncertain, so the belief evolves with time -- hence, the ``intensive margin'' of learning becomes relevant.

Another literature has studied experimentation in a voting framework. In experimentation models, a committee at each point in time chooses between a risky and a safe alternative. Each voter is uncertain about her payoff from the risky alternative. As long as the committee is choosing the risky alternative, for each voter a signal may arrive, revealing her individual preferences. In this setting, \citet{strulovici2010learning}, shows that the committee stops experimenting too early compared to the social optimum -- a conclusion that is the opposite of the result of this paper. In a related framework, \citet{messner2012option} analyse the effect of different voting rules, showing that a supermajority rule is optimal. \citet{hudja2019voting} and \citet{freer2020collective} examine a similar setting in laboratory experiments. These papers, however, study a very different type of environments: voters are ex ante identical, each voter's decision is driven by her own belief about her type, and acquiring information requires implementing the risky alternative. In my model, on the other hand, voters are ex ante heterogeneous and have opposing preferences and known factional identities; the collective decision is driven by is a \emph{common} belief about the type of the project; and acquiring information requires choosing the safe alternative, while choosing the risky option ends the game. 

More broadly, the idea that payoff consequences of collective decisions are ex ante uncertain dates back to at least \citet{fernandez1991resistance}. In that paper, a welfare-improving reform can have unanimous support when its consequences are uncertain. After approval, the identities of winners and losers are revealed, and the reform may be reversed. In my paper, on the other hand, the committee can vote to learn the outcomes of the project before approving it, and learning happens dynamically over time.

Finally, the paper adds to the literature that analyses reasons for inefficient policy choices. Prior explanations include influence from special interest groups \citep{grossman2001special,bombardini2020empirical}, inefficient contract design \citep{lewis2011procurement}, electoral considerations \citep{buisseret2018reelection}, and legislative bargaining \citep{austen2019gridlock}. This paper suggests an additional explanation, driven by the uncertainty about the distributive consequences of decisions, and about the time when this information can be revealed.


\section{Model\label{sec:Model}}

A committee consisting of factions $A$ and $B$ is considering whether to implement a project.\footnote{The equilibrium analysis treats each faction as a single player. Note that under equilibrium refinement imposed below, this is without loss of generality, as all members of a faction have the same preferences and the same information, and hence vote the same way. However, the size of each faction matters for welfare analysis of optimal decision rules.} There is continuous time $t$. At every point in time, each faction decides between voting for and against the project. Neither faction has sufficient weight to force the committee to approve the project. Hence, the project is implemented once both factions vote in favour of it. This means that any faction can delay approval by any amount of time, including, possibly, blocking it forever. When the project is approved, the game ends and payoffs (discussed below) are realised. The payoff of each faction is discounted at an exponential rate $r$.

Implementing the project imposes a cost $c\in\left(0,1\right)$ on each of the factions. At the same time, if implemented, the project generates a benefit whose value is normalised to $1$. The project has an unknown type $\theta\in\left\{ a,b\right\} $. The type of the project corresponds to the faction that receives the benefit. Thus, if a project of type $a$ is approved at time $t$, faction $A$ receives a payoff of $\left(1-c\right)e^{-rt}>0$, while faction $B$ receives a payoff of $-ce^{-rt}<0$. On the other hand, if a project of type $b$ is approved at time $t$, faction $A$ receives a payoff of $-ce^{-rt}$, while faction $B$ receives a payoff of $\left(1-c\right)e^{-rt}$. If the project is never approved, each faction receives a payoff of zero.

As long as the project is not approved, at any time a public signal may arrive and reveal its type. For a project with type $\theta$, the arrival of a signal corresponds to a jump time of a Poisson process with intensity $\lambda_{\theta}$.\footnote{Section \ref{sec:Conclusions} discusses the role of this assumption.} As long as the signal has not arrived, players are updating their beliefs about the type. Let $p_{0}$ be the common prior probability that $\theta=a$, and let $p_{t}$ be the probability that $\theta=a$ conditional on no signal arriving before time $t$. In the subsequent text, I will refer to $p_{t}$ as ``the belief''. All aspects of the game except for the type are common knowledge.

I will focus on Markov strategies with $p_{t}$ as a Markov state. For each faction $i\in\left\{ A,B\right\} $, a Markov strategy implies a set $S_{i}\subseteq\left[0,1\right]$ of beliefs, such that faction $i$ votes to approve the project if and only if $p_{t}\in S_{i}$. Thus, $S_{i}$ fully describes faction $i$'s strategy. 

Note that, once a signal arrives, $p_{t}$ remains constant at zero or at one forever. Hence, after a signal arrives, the project will never be approved, as one of the factions strictly prefers to vote against it.

If $\lambda_{a}=\lambda_{b}$, the belief does not change as the committee is delaying the decision. In that case, at any Markov equilibrium the committee either approves the project immediately, or never approves it. The former happens when both factions are in favour of the project ex ante, that is, when $p_{0}\in\left[c,1-c\right]$; otherwise, the latter happens.\footnote{For an analysis of such decisions under a more general distribution of payoffs from the project, see \citet{ginzburg2019collective}.} From now on I will focus on the more interesting case when $\lambda_{a}\neq\lambda_{b}$. Without loss of generality, I will assume that $\lambda_{a}>\lambda_{b}$.

As usual in voting games, the game has many trivial equilibria, for example, one in which each faction votes against the project at all beliefs, and neither faction deviates because it is not pivotal. For this reason, I will restrict attention to equilibria with the following characteristic: at the equilibrium a faction votes to approve the project at belief $p_{t}$ if and only if its expected payoff when the project is approved at $p_{t}$ is greater than its expected payoff from delaying the project by any (possibly infinite) amount of time before returning to playing according to the equilibrium. This refinement is similar to the standard assumption of eliminating weakly dominated strategies in static voting games.\footnote{See also \citet{strulovici2010learning} for a similar refinement in a continuous-time voting game.} Intuitively, it assumes that at each moment a player votes as if her vote was pivotal for approving or rejecting the project at that moment.

Formally, consider strategies $S_{A},S_{B}$, and take any belief $p_{t}\in\left(0,1\right)$. Let $u_{i}\left(p_{t}\right)$ be faction $i$'s expected payoff if the project is approved at belief $p_{t}$.\footnote{In the language of the literature on strategic experimentation (\citealp{keller2005strategic}), $u_{i}\left(p_{t}\right)$ is the myopic payoff of faction $i$ from adopting the project at belief $p_{t}$.} Given the strategy $S_{-i}$ of the other faction, for each $T>0$, let $V_{i}\left(p_{t},T,S_{-i}\right)$ be faction $i$'s expected payoff at belief $p_{t}$ from voting against the project for $T$ units of time before switching to voting in favour of it if no signal has arrived during that time. Strategy profile $\left(S_{A},S_{B}\right)$ constitutes an equilibrium whenever for each $i\in\left\{ A,B\right\} $ and all $p_{t}\in\left[0,1\right]$, we have $p_{t}\in S_{i}$ if and only if $u_{i}\left(p_{t}\right)\geq V_{i}\left(p_{t},T,S_{-i}\right)$ for all finite or infinite $T>0$.

\section{Results\label{sec:Results}}

\subsection{Equilibrium\label{subsec:Equilibrium}}

Consider a belief $p_{t}$. The payoffs of factions $A$ and $B$ from approving the project immediately equal

\[
u_{A}\left(p_{t}\right)=p_{t}\left(1-c\right)e^{-rt}-\left(1-p_{t}\right)ce^{-rt}=\left(p_{t}-c\right)e^{-rt},
\]
and
\[
u_{B}\left(p_{t}\right)=-p_{t}ce^{-rt}+\left(1-p_{t}\right)\left(1-c\right)e^{-rt}=\left(1-c-p_{t}\right)e^{-rt}.
\]

Note that, at a given belief, any faction can ensure a payoff of zero by voting against the project forever, as doing so will mean that the project is never adopted. Consequently, under the equilibrium refinement described above, faction $i\in\left\{ A,B\right\} $ will not vote for the project when its instantaneous payoff $u_{i}\left(p_{t}\right)$ is negative. Therefore, faction $A$ will vote against the project at all $p_{t}<c$, while faction $B$ will vote against the project at all $p_{t}>1-c$. As a consequence, if $c>\frac{1}{2}$, then at any belief at least one of the factions will vote against the project, and hence the project will never be approved. If $c\leq\frac{1}{2}$, the project can only be approved when $p_{t}\in\left[c,1-c\right]$.

As the approval is being delayed, players are updating their beliefs about the type of the project. If no signal arrives by time $t$, the belief equals
\begin{equation}
p_{t}=\frac{p_{0}e^{-\lambda_{a}t}}{p_{0}e^{-\lambda_{a}t}+\left(1-p_{0}\right)e^{-\lambda_{b}t}}=\frac{1}{1+\frac{1-p_{0}}{p_{0}}e^{\left(\lambda_{a}-\lambda_{b}\right)t}}.\label{eq: belief evolution}
\end{equation}

Note that $p_{t}$ is decreasing with $t$. Hence, with time, faction $A$ becomes increasingly more pessimistic about its payoff from the project. Consequently, it receives no benefit from learning, and has no incentive to delay voting for the project in order to learn its type. Instead, it votes for the project whenever its myopic payoff $u_{A}\left(p_{t}\right)$ from adopting the project is greater than zero -- that is, as long as $p_{t}\geq c$. The following proves this formally:
\begin{lem}
\label{lem: cutoff A, 1-sided}At any equilibrium, faction $A$ votes for the project if and only if $p_{t}\in\left[c,1\right]$.
\end{lem}
This means, in particular, that if the initial belief $p_{0}$ is below $c$, the project is never approved. Thus, the project cannot be approved if $c>\frac{1}{2}$, or if $c>p_{0}$. From now on, the paper will focus on the more interesting case when $c\leq\min\left\{ p_{0},\frac{1}{2}\right\} $.

Now consider the decision of faction $B$. This faction becomes more optimistic about the project with time. When choosing whether to vote for the project, it faces a tradeoff. On the one hand, delaying the decision makes it more likely that the project is approved at a more favourable belief -- in other words, it reduces the probability that a project of type $a$ is approved. On the other hand, delaying reduces the payoff from the project due to discounting; it may also cause type $b$ to be revealed, preventing the committee from implementing a project that benefits faction $B$. This tradeoff constitutes an optimal stopping problem for faction $B$. However, approval of the project also requires support of faction $A$, and this support will disappear once the belief falls below $c$. Hence, faction $B$ chooses the belief at which the project is approved, with the restriction that if the project is not approved when the belief reaches $c$, it will never be approved.

If without the restriction the solution to the stopping problem involves stopping at the belief that is greater than $c$, then it also optimal for faction $B$ to switch to approving the project at that belief. On the other hand, in case the unrestricted solution involves stopping at a belief below $c$, then at $p_{t}=c$ faction $B$ knows that continuing to oppose the project means that its payoff is zero. At the same time, supporting the project at belief $p_{t}=c$ gives faction $B$ an instantaneous payoff of $\left[1-2c\right]e^{-rt}$, which is positive, as $c\leq\frac{1}{2}$ by assumption. Hence, in this case faction $B$ switches its vote at $p_{t}=c$. This reasoning implies that the equilibrium is as follows:
\begin{lem}
\label{lem: stopping set 2-sided} Let $\bar{c}:=\frac{\sqrt{\lambda_{b}+r}}{\sqrt{\lambda_{a}+r}+\sqrt{\lambda_{b}+r}}<\frac{1}{2}$, and let
\[
p^{*}:=\begin{cases}
\frac{1}{1+\frac{\lambda_{a}+r}{\lambda_{b}+r}\frac{c}{1-c}}\geq c & \text{if }c\in\left(0,\bar{c}\right],\\
c & \text{if }c\in\left(\bar{c},\frac{1}{2}\right].
\end{cases}
\]

If $p_{0}>p^{*}$, then at the equilibrium the project is approved at belief $p^{*}$ and at time $t^{*}:=\frac{1}{\lambda_{a}-\lambda_{b}}\ln\left(\frac{p_{0}}{1-p_{0}}\frac{1-p^{*}}{p^{*}}\right)$, unless a signal arrives before that. If $p_{0}\leq p^{*}$, then at the equilibrium the project is approved immediately with probability one.
\end{lem}
To see the intuition, consider the case when the initial belief $p_{0}$ is greater than $p^{*}$. When $c\leq\bar{c}$, the solution of the aforementioned unrestricted stopping problem (which in this case equals $p^{*}$) is weakly greater than $c$. In this case, faction $B$ switches to supporting the project when the belief reaches that solution. On the other hand, if $c>\bar{c}$, the unrestricted solution is smaller than $c$. Then faction $B$ delays approval until the belief reaches $c$. When it does, faction $B$ votes for the project, because otherwise the project is never implemented. Finally, if $p_{0}$ is already below the optimal solution, the project is approved at the start.

\subsection{Project quality and approval chance}

We can now turn to the first of the main results, which relates the project's efficiency to its chance of being approved. Recall that a project, if implemented, imposes a cost $c$ on each faction. Thus, given the belief, a project is better for both factions when $c$ is lower. Does the outcome of the vote reflect this efficiency ranking?

If $p_{0}\leq p^{*}$, the project is approved immediately with probability one. If $p_{0}>p^{*}$, the project is approved when the belief reaches $p^{*}$, if no signal arrives before that. Hence, approval is more likely if it takes less time for the belief to reach $p^{*}$. This happens if $p^{*}$ is closer to $p_{0}$, that is, if $p^{*}$ is higher. Given the expression for $p^{*}$ in Lemma \ref{lem: stopping set 2-sided}, this implies the following result:
\begin{prop}
\label{pr: approval c}If $p_{0}\leq p^{*}$, the probability that the project is approved does not depend on $c$. Otherwise, the probability that the project is approved is decreasing in $c$ when $c<\bar{c}$, and increasing in $c$ if $c>\bar{c}$.
\end{prop}
Thus, when the cost of the project relative to its benefit is high, the probability of approval is increasing in $c$. In other words, making a project ex ante and ex post worse for both factions \emph{increases }its chance of being approved. For instance, recall the example in the Introduction of the legislature considering whether to build a railway line. Proposition \ref{pr: approval c} says that as is the cost of the railway is sufficiently high, a further increase in the cost makes the legislature is more likely to support building the railway.

The reason for this is that faction $A$ is more reluctant to support a project with a higher cost. Hence it switches from supporting to opposing the project earlier. As a result, faction $B$ will wait less before switching to support the project. By the above reasoning, this implies a higher probability of approval.

Figure \ref{fig:Pr(app), 2-sided} shows the probability of approval as a function of the project's cost $c$. If $c$ is very low, then $p^{*}$ is higher than the initial belief $p_{0}$, so the project is approved immediately. For moderately high values of $c$, the probability of approval is decreasing with cost. When $c$ becomes high enough, a further increase makes approval more likely.

\begin{figure}
\begin{centering}
\begin{tikzpicture}[yscale=5, xscale=20]

\draw[->] (0,0) -- (0.55,0) node[right] {$c$};       
\draw[->] (0,0) -- (0,1.1); 

\draw (0,0) node[left,below]{0};
\draw (0,1) node[left]{1};
\draw (1,0) node[below]{1};


\draw[red,very thick,domain=0:2/29][samples=100] plot(\x, {1});
\draw[red,very thick,domain=2/29:1/4][samples=100] plot(\x, {(0.6^(-3/32))*(0.4^(35/32))*((9*\x/(1-\x))^(-3/32)+(9*\x/(1-\x))^(-35/32))});
\draw[red,very thick,domain=1/4:0.5][samples=100] plot(\x, {(0.6^(-3/32))*(0.4^(35/32))*(((1-\x)/\x)^(-3/32)+((1-\x)/\x)^(-35/32))});

\draw[dashed] (2/29,1) -- (2/29,0) node[below]{$\frac{2}{29}$};
\draw[dashed] (1/4,0.46319093867) -- (1/4,0) node[below]{$\bar{c}=\frac{1}{4}$};
\draw[dashed] (0.5,0.77016083817) -- (0.5,0) node[below]{$\frac{1}{2}$};

\end{tikzpicture}
\par\end{centering}
\centering{}\caption{\label{fig:Pr(app), 2-sided}Probability that the project is approved as a function of $c$, for $r=1$, $\lambda_{a}=35$, $\lambda_{b}=3$, and $p_{0}=0.6$.}
\end{figure}
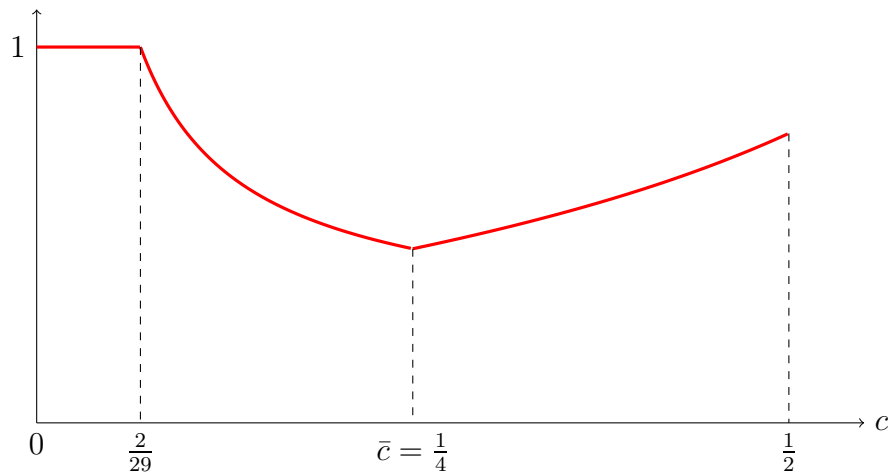

The range of $c$ for which the probability of approval is increasing in $c$ depends on the discount rate and the rates at which information is revealed. In particular, if the committee is patient and only type $A$ can be conclusively revealed, a less efficient project always has a higher chance of being approved, as the next result shows:
\begin{cor}
\label{cor:one-sided}If $r=0$ and $\lambda_{b}=0$, an increase in $c$ always increases the probability of approval.
\end{cor}
Returning to the example in the Introduction, suppose that region $B$ has better supply of skilled labour, but is politically less stable. The firm has tentatively decided to build the factory in region $B$ in case the railway is built. But an outbreak of political instability may happen that region at some point, in which case the firm will instead build it in region $A$. If members of the legislature are patient, a costlier railway project always has a higher chance of being approved.

\subsection{Prior belief, approval chance, and payoffs}

In addition to efficiency, another factor that characterises the project is the ex ante distribution of payoffs in case of approval. This distribution is captured by $p_{0}$, the probability that the project benefits faction $A$. When $p_{0}\leq p^{*}$, the project is approved immediately, so a change in $p_{0}$ does not generally change the outcome. On the other hand, when $p_{0}>p^{*}$, the project is more likely to be approved if no signal arrives during the time it takes the belief to evolve from $p_{0}$ to $p^{*}$. Thus, the project is more likely to be approved when $p_{0}$ is lower. Furthermore, the project, if approved, gives each faction a weakly positive payoff -- hence, the payoff of each faction is weakly higher if $p_{0}$ is lower. Formally, we have the following result:
\begin{prop}
\label{pr: approval p0}Suppose $p_{0}>p^{*}$. The probability that the project is approved is decreasing in $p_{0}$. The expected payoff of faction $A$ is strictly increasing in $p_{0}$ if $c<\bar{c}$ and is constant in $p_{0}$ if $c\geq\bar{c}$. The expected payoff of faction $B$ is strictly increasing in $p_{0}$ if $c<\frac{1}{2}$ and is constant in $p_{0}$ if $c=\frac{1}{2}$.
\end{prop}
Recall that approval requires consent of both factions. Nevertheless, even though the two factions seemingly have the same amount of power, Proposition \ref{pr: approval p0} implies that the probability of approval is not independent of the expected distribution of payoffs. Instead, the project is more likely to be approved if it is ex ante less likely to benefit faction $A$. The reason is that even though consensus is required, at the equilibrium the project is approved at the moment when faction $B$ switches its decision. Hence, faction $B$ is the one who has the power to decide when the project is approved.

Moreover, a project that is ex ante worse for faction $A$ weakly increases the expected payoff not only of faction $B$, but also of faction $A$, and the latter relationship is strict if $c<\bar{c}$. The reason is that if $c<\bar{c}$, faction $A$ receives a strictly positive expected payoff if the project is approved. Because lower $p_{0}$ means higher chance of approval, it implies a higher expected payoff for faction $A$. In contrast, if $c\geq\bar{c}$, approval of the project, if it happens at all, takes place at a belief $p^{*}=c$, when faction $A$ is indifferent between approving and rejecting the project -- hence, its ex ante expected payoff is zero, regardless of $p_{0}$.

This result implies that if members of faction $A$ can choose which project will be considered by the committee, they will, ceteris paribus, choose a project that is ex ante less likely to benefit their own faction.

\subsection{Optimal amount of learning and the efficient deadline}

If $p_{0}\leq p^{*}$, at the equilibrium the proposal is approved without any delay. If $p_{0}>p^{*}$, the committee acquires some information before making its decision. A longer delay means that more information is acquired. Is the equilibrium amount of information acquisition too large, too small, or optimal?

Part of the reason the question matters is that this amount can be modified by changing the decision-making procedure. In many decision-making bodies there is a minimum waiting time between submission of a proposal and a final decision on it. For example, parliaments often require several readings to approve a law, with some delay between them. Such arrangements impose a minimum amount of learning before a bill can be adopted. Conversely, some decision-making procedures impose a deadline after which the proposal cannot be approved -- this has the effect of limiting the amount of information that is acquired. Under the utilitarian welfare criterion, when are such rules optimal?

Suppose the share of faction $A$'s members is $\alpha$, and the share of faction $B$'s members is $1-\alpha$. Thus, $\alpha$ is the weight of faction $A$'s utility in the utilitarian welfare function. The effect of decision-making rules depends on $\alpha$, as the following result shows:
\begin{prop}
\label{pr: deadline} If $p_{0}\leq p^{*}$, not imposing a decision rule is weakly optimal. If $p_{0}>p^{*}$, there exists $\bar{\alpha}\in\left(0,c\right)$ such that not imposing a decision rule is optimal if $\alpha<\bar{\alpha}$ and $c>\bar{c}$. In all other cases, a deadline is optimal.
\end{prop}
Hence, minimum delay is never socially optimal. A deadline has no effect if $p_{0}\leq p^{*}$, because in this case the committee does not acquire information anyway. Otherwise, a deadline is optimal, unless the share of faction $A$ is small \emph{and} the cost of the project is large. In other words, the amount of information that the committee acquires is never too small, and in many cases is too large.

Intuitively, increasing information acquisition has two effects. First, it increases the probability that the type of the project is revealed. As a consequence, approval becomes less likely. However, at the equilibrium the project can only be approved at a belief at which both factions receive weakly positive expected payoffs (and, unless $c=\frac{1}{2}$, one faction receives a strictly positive payoff). Consequently, this effect reduces aggregate welfare. Second, more information acquisition causes the project to be approved when the belief is lower, which benefits faction $B$ at the expense of faction $A$. However, at the equilibrium faction $B$ already has the power to effectively decide when the project is approved -- thus, the second effect does not increase its payoff. Because of this, increasing the amount of learning by imposing a minimum delay does not increase welfare.

On the other hand, reducing learning by imposing a deadline reverses the two effects: it increases aggregate welfare while redistributing it from faction $B$ to faction $A$. The first effect is relatively strong when $c$ is sufficiently small, while the second effect is strong when $\alpha$ is large. Hence, unless both $c$ is high \emph{and} $\alpha$ is low, it is strictly better to force the committee to acquire less information -- thus, a deadline strictly increases welfare.

\subsection{The Role of the Poisson News Arrival Process}

One of the features of the model is that the uncertainty is revealed via a jump in the Poisson process. While this assumption is quite specific, it can easily be relaxed. Suppose, for example, that the time at which type $\theta\in\left\{ a,b\right\} $ is revealed follows cumulative distribution $F_{\theta}$. If $\frac{1-F_{a}\left(t\right)}{1-F_{b}\left(t\right)}$ is monotone decreasing, the probability that the project favours faction $A$ is decreasing with time. Hence, as before, faction $A$ does not benefit from learning. Then it is easy to show that its equilibrium strategy is the same as in Lemma \ref{lem: cutoff A, 1-sided}: it votes in favour of the project until its instantaneous expected payoff from adopting it becomes negative. The proof is similar to that of Lemma \ref{lem: cutoff A, 1-sided}.

Faction $B$ then faces a similar tradeoff as in the baseline model: delaying enables it to adopt the project at a more favourable belief, but runs the risk of revealing type $B$, and reduces the expected payoff via discounting. Then if $r$ is low, and if type $B$ is unlikely to be revealed -- faction $B$ will delay approval of the project until faction $A$ is about to switch to opposing it. Increased cost $c$ will cause this to happen earlier, and hence will make approval more likely ex ante. Corollary \ref{cor:one-sided} thus continues to hold, as does the logic of Proposition \ref{pr: approval p0}. Therefore, the results remain the same in spirit, even if a closed-form solution for the strategy of faction $B$ cannot easily be derived. On the other hand, if the belief is not monotone, the logic of the model generally does not hold.

\section{Discussion and Conclusions\label{sec:Conclusions}}

Collective decisions that involve uncertainty are common. This paper has looked at settings in which the uncertainty concerns distributive consequences of a decision, and is resolved at an uncertain time. In such situations, collective decisions can be inefficient in several aspects. First, a proposal that is worse for everyone is more likely to be approved. Second, some members may prefer to make a proposal that is worse for their side, because such a proposal is more likely to be approved. Third, the equilibrium amount of information acquisition is inefficiently large.

Future research may look at combining the features of this model with those of the prior literature on collective learning. Much of the latter studied settings in which committee members agree about the optimal decision in each state but differ in the intensity of preferences. On the other hand, this paper looks at committees whose members have opposing preferences, but within each faction preferences are homogeneous. Subsequent research can combine the two features, considering committees in which members are divided into factions, and members of the same faction differ in the intensity of preferences.

\section*{Appendix\label{sec:Appendix}}

\paragraph{Proof of Lemma \ref{lem: cutoff A, 1-sided}.}

If $p_{t}<c$, then for any $S_{B}$, $u_{A}\left(p_{t}\right)<V_{A}\left(p_{t},\infty,S_{B}\right)=0$, where $V_{A}\left(p_{t},\infty,S_{B}\right)$ is faction $A$'s payoff from voting against the project forever. Hence, any $p_{t}<c$ does not belong to $S_{A}$.

Consider now any $p_{t}\geq c$, and fix $S_{B}$. Suppose faction $A$ votes against the project for $T$ units of time before voting in favour of it. Doing this implies a (possibly infinite) time $\tau\geq T$ such that the project is approved at time $\tau$ if no signal arrives by then. The probability that no signal arrives by the time $\tau$ equals $e^{-\lambda_{a}\tau}$ for a project of type $a$, and $e^{-\lambda_{b}\tau}$ for type $b$. Hence, for any $T>0$ and any $S_{B}$, we have
\begin{align*}
V_{A}\left(p_{t},T,S_{B}\right)= & \left[p_{t}\left(1-c\right)e^{-\lambda_{a}\tau}-\left(1-p_{t}\right)e^{-\lambda_{b}\tau}c\right]e^{-r\left(t+\tau\right)}\\
< & \left[p_{t}\left(1-c\right)e^{-\lambda_{b}\tau}-\left(1-p_{t}\right)e^{-\lambda_{b}\tau}c\right]e^{-r\left(t+\tau\right)}\\
= & \left[p_{t}\left(1-c\right)-\left(1-p_{t}\right)c\right]e^{-\lambda_{b}\tau-r\left(t+\tau\right)}\\
= & u_{A}\left(p_{t}\right)e^{-\lambda_{b}\tau-r\left(t+\tau\right)}\\
\leq & u_{A}\left(p_{t}\right),
\end{align*}
where the first inequality follows from the fact that $\lambda_{a}>\lambda_{b}$, and the last inequality -- from the fact that $u_{A}\left(p_{t}\right)\geq0$ for all $p_{t}\geq c$. Hence, if $p_{t}\geq c$, then $u_{A}\left(p_{t}\right)>V_{A}\left(p_{t},T,S_{B}\right)$ for all $T>0$, and thus $p_{t}\in S_{A}$.\qed

\paragraph{Proof of Lemma \ref{lem: stopping set 2-sided}.}

Let $t$ be the time at which faction $B$ switches to voting for the project. Let $W_{B}\left(t\right)$ be the expected payoff of faction $B$ as a function of $t$. Then faction $B$ solves
\begin{equation}
\max_{t}W_{B}\left(t\right)\text{ subject to }p_{t}\geq c.\label{eq:max Wt}
\end{equation}

Note that $W_{B}\left(t\right)=-p_{0}ce^{-\left(\lambda_{a}+r\right)t}+\left(1-p_{0}\right)\left(1-c\right)e^{-\left(\lambda_{b}+r\right)t}$, and
\[
\frac{\partial W_{B}\left(t\right)}{\partial t}=\left(\lambda_{a}+r\right)p_{0}ce^{-\left(\lambda_{a}+r\right)t}-\left(\lambda_{b}+r\right)\left(1-p_{0}\right)\left(1-c\right)e^{-\left(\lambda_{b}+r\right)t}.
\]

This derivative is positive if and only if 
\begin{align}
 & \left(\lambda_{b}+r\right)\left(1-p_{0}\right)\left(1-c\right)e^{-\left(\lambda_{b}+r\right)t}<\left(\lambda_{a}+r\right)p_{0}ce^{-\left(\lambda_{a}+r\right)t}\nonumber \\
\iff & e^{\left(\lambda_{a}-\lambda_{b}\right)t}<\frac{\lambda_{a}+r}{\lambda_{b}+r}\frac{p_{0}}{1-p_{0}}\frac{c}{1-c}.\label{dW-dt>0}
\end{align}

Suppose that $p_{0}\leq p^{*}$; note that this can only hold together with $c\leq\min\left\{ p_{0},\frac{1}{2}\right\} $ when $p^{*}\geq c$, that is, when $p^{*}=\frac{1}{1+\frac{\lambda_{a}+r}{\lambda_{b}+r}\frac{c}{1-c}}$. Then we have
\begin{align*}
 & p_{0}\leq\frac{1}{1+\frac{\lambda_{a}+r}{\lambda_{b}+r}\frac{c}{1-c}}\\
\iff & \frac{p_{0}}{1-p_{0}}\leq\frac{\lambda_{b}+r}{\lambda_{a}+r}\frac{1-c}{c}\\
\iff & \frac{\lambda_{a}+r}{\lambda_{b}+r}\frac{p_{0}}{1-p_{0}}\frac{c}{1-c}\leq1,
\end{align*}
and hence (\ref{dW-dt>0}) cannot hold, because $\lambda_{a}>\lambda_{b}$ implies that $e^{\left(\lambda_{a}-\lambda_{b}\right)t}>1$. Thus, $W_{B}\left(t\right)$ is decreasing in $t$ for all $t\geq0$, and hence the optimal $t$ equals zero, so the project is approved immediately.

Suppose instead that $p_{0}>p^{*}$. Then (\ref{dW-dt>0}) holds if and only if
\[
t<\frac{1}{\lambda_{a}-\lambda_{b}}\ln\left(\frac{\lambda_{a}+r}{\lambda_{b}+r}\frac{p_{0}}{1-p_{0}}\frac{c}{1-c}\right):=\hat{t}.
\]

Thus, $\hat{t}$ is the solution to the unrestricted stopping problem. At time $\hat{t}$, the belief equals $p_{\hat{t}}=\frac{1}{1+\frac{\lambda_{a}+r}{\lambda_{b}+r}\frac{c}{1-c}}$.

If $c\leq\frac{\sqrt{\lambda_{b}+r}}{\sqrt{\lambda_{a}+r}+\sqrt{\lambda_{b}+r}}$, then $\frac{\lambda_{a}+r}{\lambda_{b}+r}\leq\left(\frac{1-c}{c}\right)^{2}$, so $p_{\hat{t}}\ge c$, and the constraint in (\ref{eq:max Wt}) is satisfied. Then faction $B$ switches to voting for the project at $t^{*}=\hat{t}$, when the belief equals $p^{*}=p_{\hat{t}}\ge c$.

If $c>\frac{\sqrt{\lambda_{b}+r}}{\sqrt{\lambda_{a}+r}+\sqrt{\lambda_{b}+r}}$, then $\frac{\lambda_{a}+r}{\lambda_{b}+r}>\left(\frac{1-c}{c}\right)^{2}$, so $p_{\hat{t}}<c$. Then $\frac{\partial W_{B}\left(t\right)}{\partial t}>0$ for all $t$ such that $p_{t}\geq c$, and we have a corner solution at which faction $B$ switches to voting for the project at a belief $p^{*}=c$. 

Finally, the expression for $t^{*}$ follows from (\ref{eq: belief evolution}).\qed

\paragraph{Proof of Proposition \ref{pr: approval c}.}

If $p_{0}\leq p^{*}$, by Lemma \ref{lem: stopping set 2-sided} the project is approved immediately with probability one regardless of $c$. If $p_{0}>p^{*}$, then the project is approved at time $t^{*}=\frac{1}{\lambda_{a}-\lambda_{b}}\ln\left(\frac{p_{0}}{1-p_{0}}\frac{1-p^{*}}{p^{*}}\right)$ if no signal arrives before that. The ex ante probability of approval then equals

\begin{align*}
 & p_{0}e^{-\lambda_{a}t^{*}}+\left(1-p_{0}\right)e^{-\lambda_{b}t^{*}}\\
= & p_{0}\left(\frac{p_{0}}{1-p_{0}}\frac{1-p^{*}}{p^{*}}\right)^{-\frac{\lambda_{a}}{\lambda_{a}-\lambda_{b}}}+\left(1-p_{0}\right)\left(\frac{p_{0}}{1-p_{0}}\frac{1-p^{*}}{p^{*}}\right)^{-\frac{\lambda_{b}}{\lambda_{a}-\lambda_{b}}}\\
= & p_{0}^{-\frac{\lambda_{b}}{\lambda_{a}-\lambda_{b}}}\left(1-p_{0}\right)^{\frac{\lambda_{a}}{\lambda_{a}-\lambda_{b}}}\left[\left(\frac{1-p^{*}}{p^{*}}\right)^{-\frac{\lambda_{a}}{\lambda_{a}-\lambda_{b}}}+\left(\frac{1-p^{*}}{p^{*}}\right)^{-\frac{\lambda_{b}}{\lambda_{a}-\lambda_{b}}}\right].
\end{align*}

The above expression is increasing in $p^{*}$. By Lemma \ref{lem: stopping set 2-sided}, $p^{*}$ is decreasing in $c$ if $c\leq\bar{c}$, and increasing in $c$ otherwise. This implies the result.\qed

\paragraph{Proof of Corollary \ref{cor:one-sided}. }

By Lemma \ref{lem: stopping set 2-sided}, if $\lambda_{b}=r=0$, then $\bar{c}=0$. Hence, by Proposition \ref{pr: approval c}, the probability of approval is increasing in $c$ for all $c\geq0$.\qed

\paragraph{Proof of Proposition \ref{pr: approval p0}.}

If $p_{0}\leq p^{*}$, the project is approved immediately with probability one regardless of $p_{0}$. If $p_{0}>p^{*}$, then Lemma \ref{lem: stopping set 2-sided}, the project is approved at time $t^{*}=\frac{1}{\lambda_{a}-\lambda_{b}}\ln\left(\frac{p_{0}}{1-p_{0}}\frac{1-p^{*}}{p^{*}}\right)$ if no signal arrives before that. The ex ante probability of approval then equals

\begin{align*}
 & p_{0}e^{-\lambda_{a}t^{*}}+\left(1-p_{0}\right)e^{-\lambda_{b}t^{*}}\\
= & p_{0}\left(\frac{p_{0}}{1-p_{0}}\frac{1-p^{*}}{p^{*}}\right)^{-\frac{\lambda_{a}}{\lambda_{a}-\lambda_{b}}}+\left(1-p_{0}\right)\left(\frac{p_{0}}{1-p_{0}}\frac{1-p^{*}}{p^{*}}\right)^{-\frac{\lambda_{b}}{\lambda_{a}-\lambda_{b}}}\\
= & p_{0}^{-\frac{\lambda_{b}}{\lambda_{a}-\lambda_{b}}}\left(1-p_{0}\right)^{\frac{\lambda_{a}}{\lambda_{a}-\lambda_{b}}}\left[\left(\frac{1-p^{*}}{p^{*}}\right)^{-\frac{\lambda_{a}}{\lambda_{a}-\lambda_{b}}}+\left(\frac{1-p^{*}}{p^{*}}\right)^{-\frac{\lambda_{b}}{\lambda_{a}-\lambda_{b}}}\right]
\end{align*}
which is strictly decreasing in $p_{0}$.

To see the effect of $p_{0}$ on payoffs, note that the expected payoff of faction $A$ equals
\begin{align*}
 & p_{0}e^{-\lambda_{a}t^{*}}\left(1-c\right)-\left(1-p_{0}\right)e^{-\lambda_{b}t^{*}}c\\
= & p_{0}^{-\frac{\lambda_{b}}{\lambda_{a}-\lambda_{b}}}\left(1-p_{0}\right)^{\frac{\lambda_{a}}{\lambda_{a}-\lambda_{b}}}\left[\left(\frac{1-p^{*}}{p^{*}}\right)^{-\frac{\lambda_{a}}{\lambda_{a}-\lambda_{b}}}\left(1-c\right)-\left(\frac{1-p^{*}}{p^{*}}\right)^{-\frac{\lambda_{b}}{\lambda_{a}-\lambda_{b}}}c\right].
\end{align*}

This is decreasing in $p_{0}$ if and only if 
\begin{align}
 & \left(\frac{1-p^{*}}{p^{*}}\right)^{-\frac{\lambda_{a}}{\lambda_{a}-\lambda_{b}}}\left(1-c\right)-\left(\frac{1-p^{*}}{p^{*}}\right)^{-\frac{\lambda_{b}}{\lambda_{a}-\lambda_{b}}}c\geq0\nonumber \\
\iff & \frac{1-c}{c}\geq\frac{1-p^{*}}{p^{*}}.\label{eq:uA dec in p0}
\end{align}

If $c<\bar{c}$, then by Lemma \ref{lem: stopping set 2-sided}, $p^{*}>c$, so (\ref{eq:uA dec in p0}) holds with a strict inequality. If $c=\bar{c}$ or if $c>\bar{c}$, then by Lemma \ref{lem: stopping set 2-sided}, $p^{*}=c$, so (\ref{eq:uA dec in p0}) holds with equality.

On the other hand, the expected payoff of faction $B$ equals
\begin{align*}
 & -p_{0}e^{-\lambda_{a}t^{*}}c+\left(1-p_{0}\right)e^{-\lambda_{b}t^{*}}\left(1-c\right)\\
= & p_{0}^{-\frac{\lambda_{b}}{\lambda_{a}-\lambda_{b}}}\left(1-p_{0}\right)^{\frac{\lambda_{a}}{\lambda_{a}-\lambda_{b}}}\left[-\left(\frac{1-p^{*}}{p^{*}}\right)^{-\frac{\lambda_{a}}{\lambda_{a}-\lambda_{b}}}c+\left(\frac{1-p^{*}}{p^{*}}\right)^{-\frac{\lambda_{b}}{\lambda_{a}-\lambda_{b}}}\left(1-c\right)\right].
\end{align*}

This is decreasing in $p_{0}$ if and only if 
\begin{align}
 & -\left(\frac{1-p^{*}}{p^{*}}\right)^{-\frac{\lambda_{a}}{\lambda_{a}-\lambda_{b}}}c+\left(\frac{1-p^{*}}{p^{*}}\right)^{-\frac{\lambda_{b}}{\lambda_{a}-\lambda_{b}}}\left(1-c\right)\geq0\nonumber \\
\iff & \frac{1-p^{*}}{p^{*}}\geq\frac{c}{1-c}.\label{eq:uB dec in p0}
\end{align}

If $c\leq\bar{c}$, then by Lemma \ref{lem: stopping set 2-sided}, $p^{*}=\frac{1}{1+\frac{\lambda_{a}+r}{\lambda_{b}+r}\frac{c}{1-c}}$, so (\ref{eq:uB dec in p0}) is equivalent to $\frac{\lambda_{a}+r}{\lambda_{b}+r}\frac{c}{1-c}\geq\frac{c}{1-c}$, which holds with a strict inequality since $\lambda_{a}>\lambda_{b}$. If $c>\bar{c}$, then by Lemma \ref{lem: stopping set 2-sided}, $p^{*}=c$, so (\ref{eq:uB dec in p0}) is equivalent to $\frac{1-c}{c}\geq\frac{c}{1-c}$, which holds with a strict inequality if $c<\frac{1}{2}$, and with equality if $c=\frac{1}{2}$.\qed

\paragraph{Proof of Proposition \ref{pr: deadline}.}

If the project is approved, both factions receive a weakly positive payoff, and at least one faction receives a strictly positive payoff. Thus, the project is ex ante welfare improving, and any deadline or a minimum waiting time that ensures that the project will not be approved reduces social welfare. Hence, we can without loss of generality focus on decision rules under which the project is approved with positive probability.

Such a decision rule implies a time $T$ such that the project is approved at $T$ if no signal arrives by then. A project of type $\theta\in\left\{ a,b\right\} $ is then approved with ex ante probability $e^{-\lambda_{\theta}T}$. Let $W\left(T\right)$ denote the utilitarian social welfare for a decision rule with a given $T$. It is given by
\begin{align*}
W\left(T\right)= & p_{0}\left[\alpha\left(1-c\right)-\left(1-\alpha\right)c\right]e^{-\left(\lambda_{a}+r\right)T}+\left(1-p_{0}\right)\left[\left(1-\alpha\right)\left(1-c\right)-\alpha c\right]e^{-\left(\lambda_{b}+r\right)T}\\
= & p_{0}\left(\alpha-c\right)e^{-\left(\lambda_{a}+r\right)T}+\left(1-p_{0}\right)\left(1-\alpha-c\right)e^{-\left(\lambda_{b}+r\right)T}.
\end{align*}

For the project to be approved at time $T$, the belief $p_{T}$ at that time must satisfy $p_{T}\in\left[c,1-c\right]$. Furthermore, we must have $p_{T}\leq p_{0}$. Hence, the set of feasible values of $T$ is given by $p_{T}\in\left[c,\min\left\{ 1-c,p_{0}\right\} \right].$ Using (\ref{eq: belief evolution}), this is equivalent to 

\[
e^{\left(\lambda_{a}-\lambda_{b}\right)T}\in\left[\max\left\{ \frac{p_{0}}{1-p_{0}}\frac{c}{1-c},1\right\} ,\frac{p_{0}}{1-p_{0}}\frac{1-c}{c}\right].
\]

Hence, the optimal $T$ is a solution to
\[
\max_{T}W\left(T\right)\text{ subject to \ensuremath{e^{\left(\lambda_{a}-\lambda_{b}\right)T}\in\left[\max\left\{  \frac{p_{0}}{1-p_{0}}\frac{c}{1-c},1\right\}  ,\frac{p_{0}}{1-p_{0}}\frac{1-c}{c}\right]}.}
\]

Differentiating, we obtain
\[
\frac{\partial W\left(T\right)}{\partial T}=\left(\lambda_{a}+r\right)p_{0}\left(c-\alpha\right)e^{-\left(\lambda_{a}+r\right)T}-\left(\lambda_{b}+r\right)\left(1-p_{0}\right)\left(1-\alpha-c\right)e^{-\left(\lambda_{b}+r\right)T}.
\]

Recall that by Lemma \ref{lem: stopping set 2-sided}, without intervention the project is approved at $t=0$ if $p_{0}\leq p^{*}$, and at time $t^{*}>0$ if $p_{0}>p^{*}$. We have four cases:

\subparagraph{Case 1: $\alpha\in\left[c,1-c\right]$.}

Then $\frac{\partial W\left(T\right)}{\partial T}<0$ for all $T$. Hence, the smallest possible $T$ is optimal. Therefore, a deadline is optimal if $p_{0}>p^{*}$, and no intervention is needed if $p_{0}\leq p^{*}$. 

\subparagraph{Case 2: $\alpha>1-c\protect\geq c$.}

Then $\frac{\partial W\left(T\right)}{\partial T}>0$ if and only if 
\[
e^{\left(\lambda_{a}-\lambda_{b}\right)T}>\frac{\left(\lambda_{a}+r\right)p_{0}\left(\alpha-c\right)}{\left(\lambda_{b}+r\right)\left(1-p_{0}\right)\left(\alpha-\left[1-c\right]\right)}.
\]

Note that for any $c\leq\frac{1}{2}$, the right-hand side of the above inequality is monotone decreasing in $\alpha$ for all $\alpha\in\left(1-c,1\right)$, and reaches $\frac{\lambda_{a}+r}{\lambda_{b}+r}\frac{p_{0}}{1-p_{0}}\frac{1-c}{c}>\frac{p_{0}}{1-p_{0}}\frac{1-c}{c}$ when $\alpha=1$. Hence, $\frac{\left(\lambda_{a}+r\right)p_{0}\left(\alpha-c\right)}{\left(\lambda_{b}+r\right)\left(1-p_{0}\right)\left(\alpha-\left[1-c\right]\right)}>\frac{p_{0}}{1-p_{0}}\frac{1-c}{c}$ for all $\alpha\leq1$, so $\frac{\partial W\left(T\right)}{\partial T}<0$ for all $e^{\left(\lambda_{a}-\lambda_{b}\right)T}\in\left[\max\left\{ \frac{p_{0}}{1-p_{0}}\frac{c}{1-c},1\right\} ,\frac{p_{0}}{1-p_{0}}\frac{1-c}{c}\right]$. Therefore, a deadline is optimal if $p_{0}>p^{*}$, and no intervention is needed if $p_{0}\leq p^{*}$. 

\subparagraph*{Case 3: $\alpha<c\protect\leq1-c$ and $p_{0}>p^{*}$.}

Then $\frac{\partial W\left(T\right)}{\partial T}>0$ if and only if 

\begin{equation}
e^{\left(\lambda_{a}-\lambda_{b}\right)T}<\frac{\lambda_{a}+r}{\lambda_{b}+r}\frac{p_{0}}{1-p_{0}}\frac{c-\alpha}{1-c-\alpha}.\label{eq:case3 optimal T}
\end{equation}

By Lemma \ref{lem: stopping set 2-sided}, at the equilibrium the project is approved (if no signal arrives) at a time $t^{*}$ given by
\[
e^{\left(\lambda_{a}-\lambda_{b}\right)t^{*}}=\frac{p_{0}}{1-p_{0}}\frac{1-p^{*}}{p^{*}}.
\]

A deadline increases welfare if $t^{*}$ is greater than $T$ as defined in (\ref{eq:case3 optimal T}), that is, if
\begin{equation}
\frac{1-p^{*}}{p^{*}}>\frac{\lambda_{a}+r}{\lambda_{b}+r}\frac{c-\alpha}{1-c-\alpha},\label{eq:case3 deadline optimal}
\end{equation}
and a minimum waiting time increases welfare if $t^{*}<T$, that is, if the sign in (\ref{eq:case3 deadline optimal}) is reversed.

By Lemma \ref{lem: stopping set 2-sided}, if $c\leq\bar{c}$, then $\frac{1-p^{*}}{p^{*}}=\frac{\lambda_{a}+r}{\lambda_{b}+r}\frac{c}{1-c}$, so (\ref{eq:case3 deadline optimal}) becomes
\[
\frac{c}{1-c}>\frac{c-\alpha}{1-c-\alpha},
\]
which holds for any $\alpha\in\left(0,c\right)$. Hence, in this case a minimum waiting time is not optimal, and a deadline is optimal.

On the other hand, if $c>\bar{c}$, then $p^{*}=c$. A minimum waiting time either has no effect, or prevents the project from being approved. Neither of these increases welfare. A deadline is optimal if (\ref{eq:case3 deadline optimal}) holds, that is, if and only if
\begin{equation}
\frac{1-c}{c}>\frac{\lambda_{a}+r}{\lambda_{b}+r}\frac{c-\alpha}{1-c-\alpha}.\label{eq:case3 deadline cutoff}
\end{equation}
For $\alpha=c$ this holds. For $\alpha=0$, (\ref{eq:case3 deadline cutoff}) becomes 
\begin{align*}
 & \frac{1-c}{c}>\frac{\lambda_{a}+r}{\lambda_{b}+r}\frac{c}{1-c}\\
\iff & c<\frac{\sqrt{\lambda_{b}+r}}{\sqrt{\lambda_{a}+r}+\sqrt{\lambda_{b}+r}}=\bar{c},
\end{align*}
which not hold when $c>\bar{c}$. Because the right-hand side of (\ref{eq:case3 deadline cutoff}) is monotone in $\alpha$ for $\alpha\in\left(0,c\right)$, there exists a value $\bar{\alpha}\in\left(0,c\right)$ such that when $c>\bar{c}$, a deadline is optimal if and only if $\alpha\geq\bar{\alpha}$, and no intervention is optimal otherwise.

\subparagraph*{Case 4: $\alpha<c\protect\leq1-c$ and $p_{0}\protect\leq p^{*}$.}

Then at the equilibrium the project is approved at time $t^{*}=0$, so a deadline has no effect. On the other hand, for a minimum waiting time to increase welfare, (\ref{eq:case3 optimal T}) must hold at $T=0$, that is, we must have
\begin{align}
 & \frac{\lambda_{a}+r}{\lambda_{b}+r}\frac{p_{0}}{1-p_{0}}\frac{c-\alpha}{1-c-\alpha}>1\nonumber \\
\iff & p_{0}>\frac{1}{1+\frac{\lambda_{a}+r}{\lambda_{b}+r}\frac{c-\alpha}{1-c-\alpha}}.\label{eq:case4 MWT optimal}
\end{align}

By Lemma \ref{lem: stopping set 2-sided}, either $p^{*}=c$, or $p^{*}=\frac{1}{1+\frac{\lambda_{a}+r}{\lambda_{b}+r}\frac{c}{1-c}}$. In the former case, a minimum waiting time means that the project is never approved, which does not increase welfare. In the latter case, using the fact that $p_{0}\leq p^{*}$ and that $\frac{c-\alpha}{1-c-\alpha}$ is strictly decreasing in $\alpha$ for $\alpha\in\left[0,c\right]$, we have
\[
p_{0}\leq p^{*}=\frac{1}{1+\frac{\lambda_{a}+r}{\lambda_{b}+r}\frac{c}{1-c}}<\frac{1}{1+\frac{\lambda_{a}+r}{\lambda_{b}+r}\frac{c-\alpha}{1-c-\alpha}},
\]
and hence (\ref{eq:case4 MWT optimal}) does not hold, implying that a minimum waiting time is not optimal.

\subparagraph*{Summary of the four cases.}

Neither a deadline nor a minimum waiting time is optimal if $p_{0}\leq p^{*}$; or if $c>\bar{c}$ and $\alpha<\bar{\alpha}$ for some $\bar{\alpha}\in\left(0,c\right)$. In all other cases, a deadline increases welfare.\qed

\bibliography{dividedcommittees}

\end{document}